\documentclass{svjour2}                    
\smartqed  
\usepackage{graphicx}
\begin{document}

\title{Heterogeneity and Allometric Growth of Human Collaborative Tagging Behavior
}
\subtitle{}
    
\author{Lingfei Wu         \and
        Chengjun Wang
}

\institute{L. Wu \at
              City University of Hong Kong, Kowloon, HongKong \\
              Tel.: +852-68058840\\
              \email{wlf850927@gmail.com}           
           \and
           C. Wang \at
              City University of Hong Kong, Kowloon, HongKong \\
              \email{wangchj04@gmail.com} 
}

\date{Received: date / Accepted: date}

\maketitle

\begin{abstract}
Allometric growth is found in many tagging systems online. That is, the number of new tags ($T$) is a power law function of the active population ($P$), or $T\sim P^{\gamma}$  ($\gamma\neq1$). According to previous studies, it is the heterogeneity in individual tagging behavior that gives rise to allometric growth. These studies consider the power-law distribution model with an exponent $\beta$, regarding $1/\beta$ as an index for heterogeneity. However, they did not discuss whether power-law is the only distribution that leads to allometric growth, or equivalently, whether the positive correlation between heterogeneity and allometric growth holds in systems of distributions other than power-law. In this paper, the authors systematically examine the growth pattern of systems of six different distributions, and find that both power-law distribution and log-normal distribution lead to allometric growth. Furthermore, by introducing Shannon entropy as an indicator for heterogeneity instead of $1/\beta$, the authors confirm that the positive relationship between heterogeneity and allometric growth exists in both cases of power-law and log-normal distributions.
\end{abstract}

\section{Introduction}
\label{intro}

Studies on human collaborative tagging behavior discover the allometric growth \cite{1}\cite{2}\cite{3}\cite{4} of online tagging systems. That is, the power law relationship $T\sim P^{\gamma}$  ($\gamma\neq1$) between the number of new tags ($T$) and the active population ($P$). Previous studies tend to adopt a power-law distribution model (with an exponent $\beta$ ) to explain the allometric growth\cite{1}\cite{2}\cite{3}\cite{5}. Although these models may be different from each other in details, they agree on the assumption that heterogeneity in individual tagging behavior leads to the allometric growth. Moreover, [1][5] suggest that if we regard the reciprocal of $\beta$ as an indicator of heterogeneity, the positive correlation between heterogeneity $1/\beta$ and the allometric growth rate $\gamma$ can be found. However, these studies do not discuss whether the power law distribution is the only distribution that gives rise to allometric growth, neither do they consider whether the positive correlation between heterogeneity and allometric growth also exist in systems of distributions other than power-law. Without answering these questions, it is hasty to derive the conclusion that heterogeneity leads to allometric growth. In the current study, we find that both power-law distribution and log-normal distribution lead to allometric growth. Further, we suggest that if we regard rescaled Shannon entropy as a more general indicator of heterogeneity than $1/\beta$, there is supporting evidence for the positive relationship between heterogeneity and allometric growth in both systems of power-law and log-normal distributions. As allometric growth has been widely discovered in many different social systems online \cite{4}\cite{6}\cite{7}\cite{8} and offline\cite{9}\cite{10}\cite{11}\cite{12}, our findings on the relationship between Shannon entropy and allometric growth may be useful in studying these systems.

\section{The allometric growth of social annotation and the explanatory model of power law distribution}
\label{sec:1}

\begin{figure*}
\includegraphics[scale=0.6]{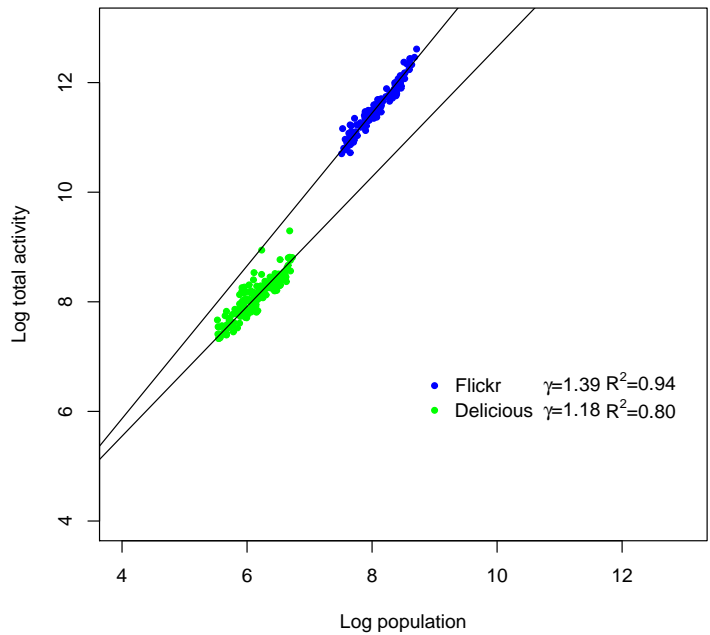}
\caption{Accelerating growth in two online tagging systems(Flickr and Delicious). Different colors indicate different types of systems. Each point is plotted by counting the number of new tags (y) and the population of active user (x) in a system in one day. The scaling exponent $\gamma$ estimated by ordinary least square (OLS) regression in the Log-Log scale plot is drawn in dark line for eye guidance. The figure is cited from \cite{1}.}
\label{fig.1}
\end{figure*}

Fig.~\ref{fig.1} shows the examples of allometric growth. To explain the growth pattern, \cite{1} proposes that if the empirical daily distribution of individual tagging activities confirms a power-law distribution invariant of system size (population), such a growth pattern can be observed. The power-law distribution model can be expressed as

\begin{equation}
\label{eq.1}
n(t) = (t/t_{max})^{-\beta}
\end{equation}
Where $t$ is the number of tags with a maximum value $t_{max}$ and $n(t)$ is the number of users that generate so many tags. \cite{1} suggests that the positive relationship between $\beta$ and $\gamma$ can be analytically derived as
\begin{eqnarray}
\label{eq.2}
\gamma = \left\{ \begin{array}{rl}
 2/\beta &\mbox{ if $1<\beta<2$ } \\
  1 &\mbox{ if $\beta\ge2$}
       \end{array} \right.
\end{eqnarray}

Eq.~\ref{eq.2} is confirmed by empirical data and numerical simulations in \cite{1}. Therefore power law distribution with an exponent smaller than 2 gives rise to allometric growth. In other words, accelerating growth usually appears in systems that are heterogeneous in resources or activities. In \cite{5} and \cite{3}, the analytical relationships between $\beta$ and $\gamma$ are slightly different from Eq.~\ref{eq.2}, but similar in terms of supporting the positive relationship between heterogeneity and allometric growth.

\section{Both power-law distributions and log-normal distributions lead to allometric growth}
\label{sec:2}

Is power-law the only distribution leading to allometric growth? Previous studies \cite{1}\cite{2}\cite{3}\cite{5} did not consider this question. To answer the question, we numerically simulate systems of six types of distributions and observe the growth rate ($\gamma$) of these systems. The detailed information of simulation is listed in Table 1. It turns out that, both the Pareto distribution with an exponent $\alpha$ between $0 \sim 1$ and the log-normal distribution give rise to allometric growth (i.e.,$\gamma > 1$). The observation that Pareto distribution leads to accelerating growth when its exponent $\alpha$ ranges from 0 to 1 is consistent with Eq.~\ref{eq.2}, since Pareto distribution (with an exponent $\alpha$) can be viewed as the cumulative distribution function of power law distribution (with an exponent $\beta$), and  $\alpha=\beta-1$.

However, it is obvious that Eq.~\ref{eq.2} can not be used to explain the allometric growth in the case of log-normal distribution. Does the relationship between heterogeneity and accelerating growth also exist in systems of non-power-law distributions (e.g., log-normal distribution)? To answer this question, we should find another indicator for heterogeneity other than $1/\beta$ . In the current research, we recommend to use Shannon entropy \cite{13} as the new indicator .

\begin{table*}
\caption{Parameter variance and growth rate ($\gamma$) in systems of different distributions}
\label{tab:1}       
\begin{tabular}{llllllll}
\hline\noalign{\smallskip}
 & Normal	& Weibull & Poisson & Gamma	& Log-normal & Pareto-1 & Pareto-2  \\\hline\noalign{\smallskip}
Parameters & $1<\mu<10$ & $1<\alpha<10$ & $0.1<\mu<10$ & $1<\alpha<10$ & $1<\mu<10$ & $1<k<10$ & $1<k<10$\\
 & $0.1<\sigma<10$ & $0.1<\beta<10$ &  & $0.1<\beta<10$ & $0.1<\sigma<10$ & $0.1<\alpha<1$ & $1<\alpha<10$\\
Mean of $\gamma$ & 1.00 & 1.00 & 1.00 & 1.00 & 1.60 & 3.02 & 1.00 \\
S.D. of $\gamma$ & 0.01 & $<0.01$ & 0.01 & $<0.01$ & 0.92 & 2.94 & 0.03 \\
N of simulations & 400 & 400 & 40 & 400 & 400 & 200 & 200 \\ \hline
\end{tabular}
\end{table*}

\section{Rescaled Shannon entropy as an indicator for heterogeneity}
\label{sec:3}

As aforementioned, we introduce the Shannon entropy $H$ as an new indicator for heterogeneity because the indicator $1/\beta$ does not apply in log-normal distribution. To test the assumption that heterogeneity leads to allometric growth, we shall examine the relationship between $H$ and $\gamma$ in both cases of power-law distribution and lognormal distribution.

According to \cite{14}, the value of entropy $H_{1}$ in power-law distribution $f(x) = C^{\beta-1}x^{-\beta}(\beta-1)$ can be calculated as
\begin{equation}
\label{eq.3}
H_{1} = k_{1}log(\frac{C}{\beta-1})+k_{2}\frac{1}{\beta-1}+k_{3}
\end{equation}

\begin{figure*}
\includegraphics[scale=0.25]{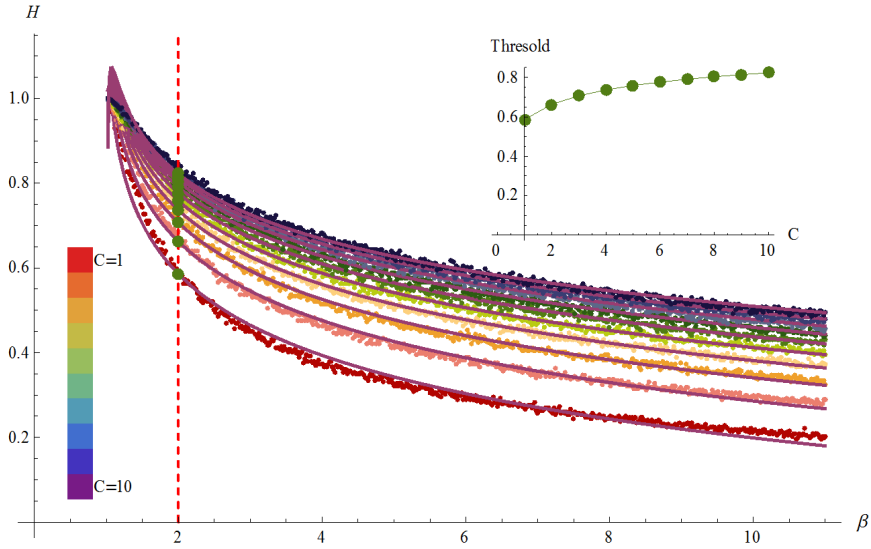}
\caption{The relation between Shannon entropy ($H$) and growth rate ($\gamma$). The result of numerical simulations (with parameters set as $1<\beta<10$ and $1<C<10$ in the power law model $f(x) = C^{\beta-1}x^{-\beta}(\beta-1)$) are plotted in dots whose color varies with $C$. The dots are fitted by function Eq.~\ref{eq.3} with OLS regression. The estimated curves are shown in purple lines. The green points with large size are the thresholds which are reached when $\beta=2$ and $C$ varies from 1 to 10. The inset shows how the value of threshold increases with the value of $C$. The minimum value of $H_{t} = 0.586$ is achieved when $C=1$ and $\beta = 2$.}
\label{fig.2}
\end{figure*}

Where $k_{1}$, $k_{2}$, and $k_{3}$ are constant coefficients to be estimated. As $H_{1}$ in Eq.~\ref{eq.3} is a system-size dependent variable, to facilitate the comparisons between entropies of systems different in size $N$, given $H_{max}=Nlog(N)$, we calculate the rescaled value of $H_{1}$  as $H=H_{1}/Nlog(N)$. In Fig.~\ref{fig.2}, We show how $H$ increases with $C$ and decreases with $\beta$ in numerical simulations. By fitting the curves in
Fig.~\ref{fig.2} with Eq.~\ref{eq.3}, we calculate the minimum value of entropy $H_{t} = 0.586$, which is achieved when $C=1$ and $\beta=2$. In another word, a system of power-law distribution will not show allometric growth unless its entropy is greater than $H_{t}=0.586$.

\begin{figure*}
\includegraphics[scale=0.8]{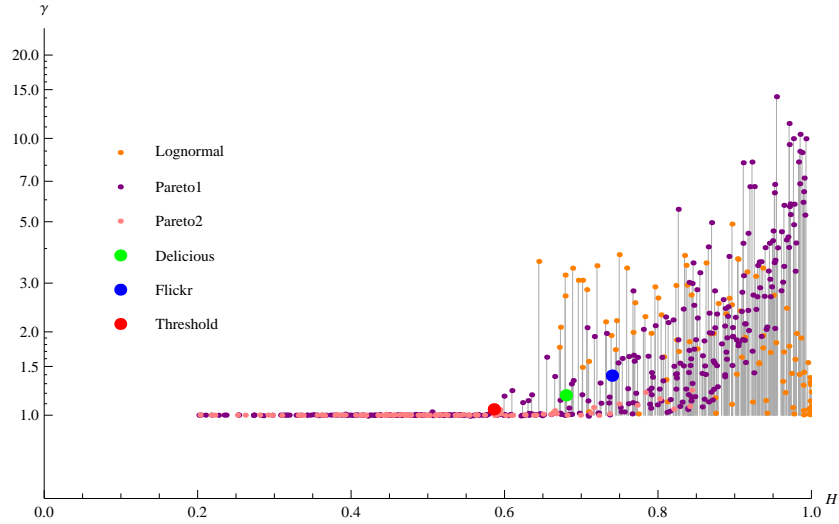}
\caption{The relationship between rescaled heterogeneity ($H$) and growth rate ($\gamma$). Both Pareto distribution with an exponent ranging from 0 to 1 (Pareto 1) and log-normal distribution lead to accelerating growth (i.e.,$\gamma>1$). Moreover, in both of the two cases, the accelerating growth would not appear until the entropy threshold $H_{t} = 0.586$ is reached.}
\label{fig.3}
\end{figure*}

In Fig.~\ref{fig.3}, we show the relationship between $H$ and $\gamma$ in simulated systems of power-law and log-normal distributions. It turns out that our finding on the threshold value of entropy $H_{t} = 0.586$ applies in both cases. We also plot the minimum entropy of two social tagging systems, Delicious and Flickr. In calculating their entropies, we set $C=1$ and substitute the empirical values of $\beta$ (Fig.~\ref{fig.1}) into Eq.~\ref{eq.3}. We find the behavior of two empirical systems is consistent with our findings. According to the results shown in Fig.~\ref{fig.3}, the positive relationship between the rescaled Shannon entropy $H$ and allometric growth rate $\gamma$ is supported both by empirical data and simulation.

\section{Discussion}
\label{sec:dis}

Previous studies uncover the allometric growth pattern of social tagging systems and explain the pattern by power-law distribution model \cite{1}\cite{2}\cite{3}\cite{5}. The authors in this paper systematically investigate the growth pattern of systems of different distributions, and find that log-normal distribution also leads to allometric growth. With regards to the principle derived from the power-law distribution model that heterogeneity leads to allometric growth \cite{1}\cite{5}, the authors propose that if we used rescaled Shannon entropy to represent the heterogeneity of individual tagging behavior, the principle holds in both cases of power-law distribution and log-normal distribution.

It is possible for the findings of this paper find their applications in the information industry. For instance, the positive relationship between heterogeneity and allometric growth indicates that to enjoy fast growth of users, webmasters should keep the threshold of their websites as low as possible to attract users as diverse as possible.

Some questions left point out the direction of future research. Why do the systems of power-law distribution and log-normal distribution, but not other distributions, show the properties of allometric growth? Can the relationship between heterogeneity and allometric growth be generalized to other online or even offline social system? These important questions call for further exploration.

\end{document}